# CARACTERISATION ELECTROMAGNETIQUE DE MATERIAUX GEOLOGIQUES EN VUE DU SUIVI DE L'HUMIDITE DES SOLS PAR RADIOMETRIE MICRO-ONDES

B. Le Crom[1,2], F. Demontoux[1], G. Ruffié[1], JP. Wigneron[2], J.P. Grant[2,3]

[1] Université de Bordeaux 1 - Laboratoire IMS-UMR 5218- 16 av Pey-Berland 33607 Pessac
[2] INRA-Unité de Bioclimatologie, BP 81, Villenave d'Ornon Cedex 33883
[3] Faculty of Earth and Life Sciences, Vrije Universiteit Amsterdam, De Boelelaam 1085.

## 1  Introduction

SMOS (Soil Moisture and Ocean Salinity), dont le lancement est prévu pour l'horizon 2007, est la seconde mission d'opportunité « Earth Explorer » à être développée dans le cadre du programme « Living Planet » de l'agence spatiale européenne (ESA) [1]. Sur la problématique du cycle de l'eau, les données acquises par SMOS permettront d'établir une carte spatiale de l'humidité des surfaces continentales et de la salinité de la couche superficielle des océans. Les applications sont multiples. Sur Terre, la rétention d'eau dans le sol joue un rôle primordial dans l'évolution climatique car l'humidité des sols représente une variable clé régulant l'échange d'eau et d'énergie thermique entre la terre et l'atmosphère. En mer, la salinité est un paramètre qui influence la circulation des masses d'eau dans les océans et entraîne la formation de phénomènes climatiques tel qu'El Niño.

Installé sur la plateforme Protéus[1], ce satellite, contenant le tout premier radiomètre interférométrique 2D, effectuera donc la 1$^e$ cartographie à l'échelle planétaire de l'humidité des sols et de la salinité des océans et ce grâce à un unique instrument de mesure capable de capture d'image des radiations micro ondes émises autour de 1.4GHz. Les hyperfréquences sont sensibles aux changements de la constante diélectrique du milieu et donc toute variation de la quantité d'eau induis des modifications des propriétés du diélectrique. Cela affecte l'émissivité, et par conséquent la température de brillance détectée par le radiomètre. Ainsi il existe une relation directe entre l'humidité du sol pour des profondeurs de 2 à 5 cm, la salinité des océans et les émissions d'origine terrestre sur la fréquence de 1.4 GHz.

En raison de la résolution spatiale du pixel de SMOS, de l'ordre de 40 Km sur 40 Km, la tache au sol de la mesure englobe généralement un grand nombre de type d'occupation du sol. Les forêts sont présentes dans une majorité des pixels en zone tropicale, boréale et tempérée. Les forêts sont des couverts relativement opaques, sur lesquels le suivi de l'humidité reste problématique. En particulier, l'effet de la litière a, jusqu'ici, été négligé.

Le but de ce travail a été d'étudier les propriétés diélectriques d'un type de litière et de terre afin d'intégrer ces valeurs à un modèle analytique multi couches de sol. L'objectif est de mettre en évidence les effets de cette strate sur le système multi couche global. Ceci permettra d'aboutir à une formulation analytique simple d'un modèle de litière qui puisse être intégré à l'algorithme de calcul de SMOS afin de recueillir des informations sur l'humidité à partir des mesures d'émissivité.

## 2  Présentation de la méthode de mesure

La première étape de cette étude a consisté à réaliser des mesures de permittivité afin de pouvoir les implanter par la suite dans les modèles numériques et analytiques de calcul d'émissivité.

Les résultats présentés concernent des échantillons prélevés le 15 février 2006 sur le site INRA du Bray (33) [2]. Ceux-ci se présentaient dans une configuration multicouche constituée par une première couche de 1 à 10 cm de litière sur le sol puis d'une couche de 10 à 30 cm de paille. Avant de procéder aux mesures les échantillons ont été pesés et séchés en étuve (terre et litière). Nous avons mesuré des permittivités faibles pour la paille, en conséquence cette étude ne tiendra pas compte de la couche de paille. Toutes les mesures ont été réalisées en guide d'onde. Cette méthode nous permet de travailler sur des volumes d'échantillons suffisamment importants pour représenter l'hétérogénéité du milieu. Les échantillons sont maintenus dans le guide à l'aide d'une bride dont le fond est constitué d'une feuille de Mylar de 100μm d'épaisseur considérée comme quasi transparente aux ondes électromagnétiques. Deux épaisseurs de brides ont été utilisées, en fonction des permittivités attendues pour chacun des matériaux. Les mesures s'effectuent par l'intermédiaire d'un analyseur de réseaux ANRITSU 37325A qui est piloté par ordinateur pour l'acquisition et le traitement des données. Les mesures sont effectuées dans une salle régulée en température. L'analyseur nous permet de mesurer les coefficients complexes $S_{21}$ et $S_{11}$ de la matrice de répartition. La détermination des paramètres électromagnétiques de l'échantillon se fait grâce à la méthode de Nicolson Ross Weir (NRW) pour les guides d'onde rectangulaires. Cette procédure de calcul de paramètres électromagnétiques d'un matériau est basée sur des mesures de réflexion et de transmission qui est bien adaptée aux analyseurs de réseaux. L'introduction de notre échantillon dans le guide produit un changement d'impédance caractéristique. L'ensemble des réflexions multiples dans le matériau est pris en compte dans les expressions des coefficients $S_{11}$ et $S_{21}$ qui sont mesurés. Ces derniers permettent alors de calculer Γ, coefficient de la première réflexion et $T$, coefficient de transmission. Un algorithme de calcul permet ensuite d'obtenir la permittivité complexe du matériau ; $\varepsilon'$ et $\varepsilon''$, sur la plage de mesure considérée, en l'occurrence [1.3GHz 1.5GHz]. La perméabilité relative du matériau sera égale à 1 car ces matériaux ne sont pas magnétiques.

---

[1] La plateforme PROTEUS (Plateforme Reconfigurable pour l'Observation, pour les Télécommunications et les Usages Scientifiques) fournit toutes les ressources nécessaires au fonctionnement du satellite dans l'espace : contrôle de la trajectoire ; puissance électrique ; communication avec la Terre.

La variation de teneur en eau de l'échantillon s'effectue par injection d'une quantité fixée d'eau distillée. Ainsi les phénomènes de diffusion de l'eau dans l'échantillon vont avoir un effet notable sur les résultats des mesures.

La Figure 1 illustre les résultats de nos mesures en fonction du temps pour une humidité volumétrique fixée à 15.54% .

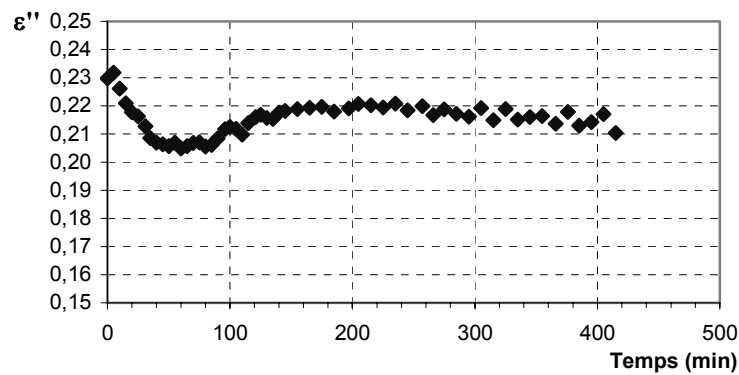

**Figure 1 : Influence, sur ε″, de la répartion de l'eau dans un échantillon de litière pour une humidité de 44.95%.**

Ces mesures permettent d'estimer l'erreur induite par le temps imposé entre chaque point.

Il a été ainsi possible de considérer qu'avec un délai de 200mn, les mesures précédentes fournissent une bonne estimation de la permittivité de la litière alors que ce délai passe à 300 mn pour la terre.

Cette étude préliminaire a permis de mettre en évidence un phénomène transitoire lors de la mesure. En effet l'absorption de l'eau n'est pas instantanée et influence donc les résultats. De plus la méthode de post traitement est prévue pour un matériau homogène, ce qui explique les fluctuations non négligeables de la permittivité lorsque le temps d'attente entre chaque mesure varie. Il est donc nécessaire de patienter suffisamment longtemps pour que l'erreur commise soit raisonnable.

Cette étude permet également de supposer que le phénomène de pluie a une influence non négligeable sur la permittivité des matériaux et donc sur l'émissivité de la structure géologique.

## 3    Présentation des mesures

Dans la nature, l'humidité volumétrique de la terre varie habituellement entre 0% et 30% (au maximum 40 à 50% à saturation). C'est pour cette raison que l'approximation induite par les mesures se limitera à ce domaine.

Rappelons les deux méthodes de calcul de l'humidité (volumétrique pour la terre et gravimétrique pour la litière) :

$$\begin{cases} SM = \dfrac{P_W^S}{P_{Dry}^S} \rho_b \\ LM = \dfrac{P_W^L}{P_{Dry}^L + P_W^L} \end{cases}$$

où :     $P_W^I$ est le poids d'eau présente dans l'échantillon où $I=S$ pour le sol et $I=L$ pour la litière ;

$P_{Dry}^I$ est le poids de l'échantillon sec ;

$\rho_b$ est la « bulk density » de la terre.

Contrairement à la terre, la litière est rarement sèche dans la nature. Son humidité gravimétrique varie plutôt entre 15% et 80%. L'étude sera donc focalisée sur ce domaine d'humidité.

Les erreurs de mesures ont plusieurs causes. Il y a celles entraînées par la pesée de l'échantillon et de la bride, celles engendrées par la méthode de mesure, par la méthode de calcul de la permittivité et enfin les erreurs résultant de la répétitivité de la méthode de mesure.

Concernant les incertitudes engendrées par la méthode de mesure, la calibration de l'analyseur de réseaux permet de corriger certaines imperfections du système mais il en subsiste comme le bruit de mesure de l'appareil, les différentes dérives, les coefficients aux différentes interfaces....

Les incertitudes liées à la méthode de calcul sont quant à elles dues aux erreurs d'arrondis réalisés lors des calculs numériques. Ces deux types d'erreurs sont difficiles à quantifier. Néanmoins, nous estimons à ±5% l'influence de cet ensemble d'incertitudes sur les valeurs des résultats de mesures.

L'erreur sur le calcul de l'humidité est principalement due aux erreurs de pesées des échantillons, pesées effectuées par l'intermédiaire d'une balance où le poids est juste au dixième près. Les masses mesurées sont donc exactes à ±0.05 g près.

Concernant les erreurs sur l'évaluation des poids, nous la prendrons égale à ±0.10% car les poids d'eau et poids secs sont calculés par différences respectivement du poids mouillé, du poids sec et du poids bride + échantillon et du poids de la bride seule.

Finalement, les calculs effectués, l'erreur de calcul de l'humidité du sol et de la litière sont respectivement de ±0.11% et de ±0.19%.

Afin de prendre en compte dans nos mesures l'hétérogénéité des milieux et la répétitivité des résultats nous avons effectué plusieurs mesures sur les échantillons en les disposant de manières différentes dans le guide.

De ces résultats nous avons pu définir l'écart type qui permet d'intégrer autour de notre résultat de mesure un domaine des permittivités possibles.

Les différentes erreurs ou variations de la mesure déterminées ou présentées ci-dessus amènent à définir non plus une courbe de type permittivité en fonction de l'humidité mais un domaine de permittivité évoluant en fonction de l'humidité du milieu considéré. Deux domaines peuvent être définis. Le premier sera appelé le domaine des permittivités possibles et le deuxième le domaine des permittivités corrigés. Le premier intègrera les fluctuations des mesures que nous avons observés (hétérogénéité des milieux) et le deuxième intègrera en plus les erreurs de mesure. Ces résultats sont présentés sur les figures ci-après ; les courbes en pointillé délimitent le domaine des permittivité possibles et les courbes pleines le domaine des permittivité corrigées.

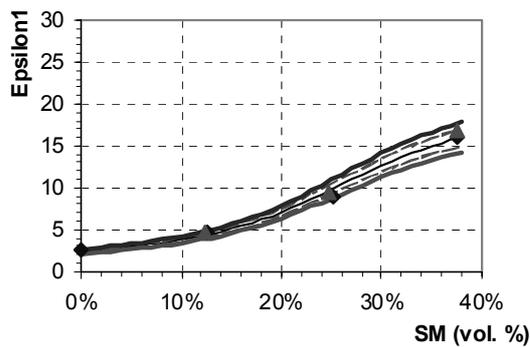

Figure 2 : Domaine de la partie réelle de la permittivité de la terre

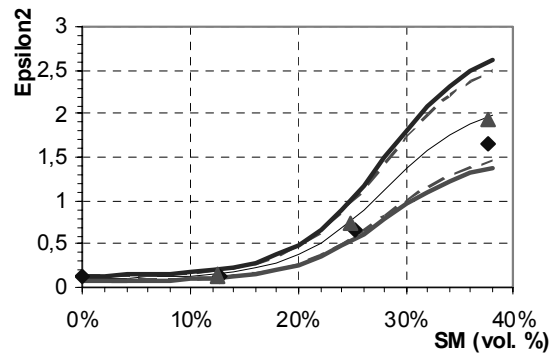

Figure 3: Domaine de la partie imaginaire de la permittivité de la terre

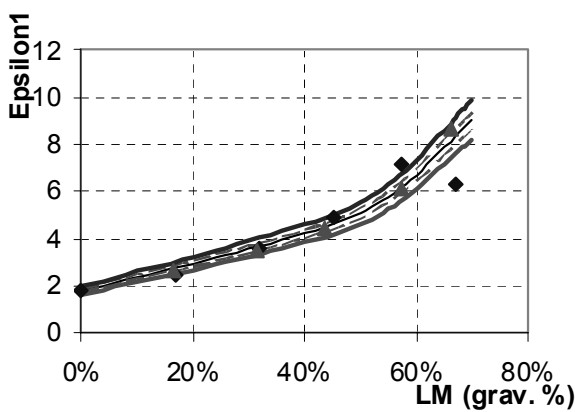

Figure 4: Domaine de la partie réelle de la permittivité de la litière

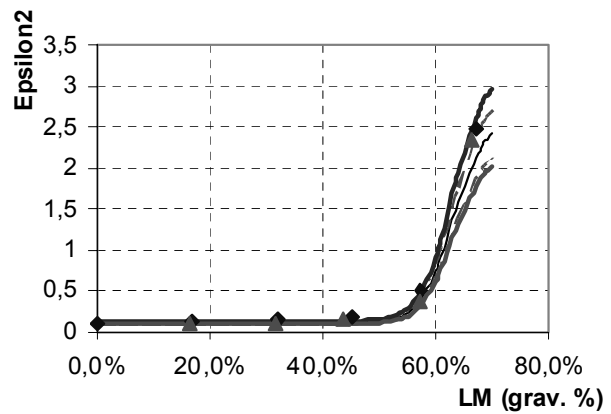

Figure 5: Domaine de la partie imaginaire de la permittivité de la litière

## 4   Exploitation des mesures

Afin d'estimer la pertinence de la précision de nos mesures nous avons intégré les domaines de permittivité mesurés dans un modèle numérique de calcul d'émissivité développé à l'aide du logiciel HFSS (High Frequency Structure Simulator) qui est un logiciel de simulation par éléments finis de la société ANSOFT. Dans ce cas précis il permet d'obtenir le coefficient $S_{11}$, puis l'émissivité, en fonction de l'humidité pour le système (multi couches ou non) étudié. Des mesures nous ont permis de définir une relation entre l'humidité du sol (SM) et celle de la litière (LM). Les résultats suivants de l'étude du système sol+litière sont présentés uniquement en fonction de SM.

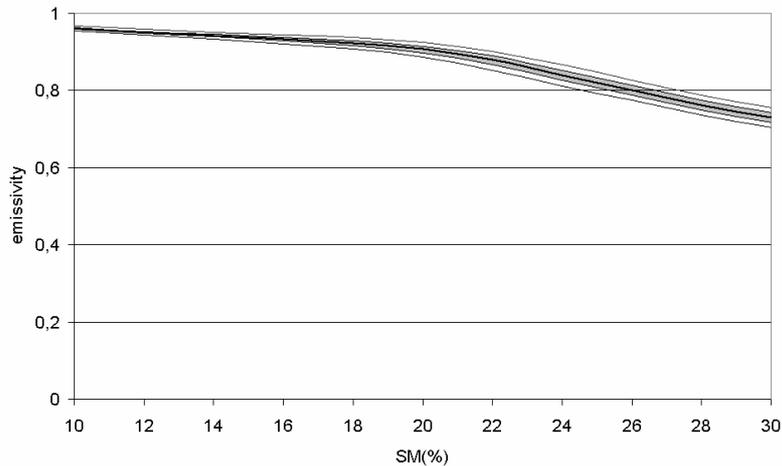

**Figure 6 : domaine d'émissivité**

Nous avons tout d'abord intégré le domaine des permittivités possible dans notre modèle HFSS bi-couche (terre + litière de 3 cm d'épaisseur). Le domaine d'émissivité calculé est représenté sur la figure ci-dessus par la zone grisée. Nous constatons que nos mesures et notre modèle nous permettent de bien appréhender la valeur d'émissivité que nous pourrions mesurer sur ce type de sol à l'aide d'un radiomètre. Nous avons par la suite intégré le domaine de permittivité corrigé. Le domaine d'émissivité obtenu est représenté non grisé sur la figure ci-dessus. Nous pouvons noter sur la figure que l'écart sur le calcul de l'émissivité, intégré par les erreurs de mesures, est au maximum de 0.01 pour une émissivité de 0.71 soit une erreur de 1.41%.

## 5   Conclusion

À la vue de nos résultats, nous constatons que notre protocole de la mesure nous permet d'évaluer correctement les caractéristiques électromagnétiques des milieux géologiques complexes tels que la litière. Un prolongement de ce travail consistera à étudier l'évolution de ces propriétés pendant un an. Ceci permettra de prendre en compte de nouveaux facteurs tels que la constitution de la litière ou de la température (la terre froide en hiver et la terre très sèche en été…). Dans cette étude nous avons considéré une couche uniforme de litière de 3 centimètres d'épaisseur et nous avons constaté que cette couche influence de manière significative l'émissivité du système de sol/litière. En réalité cette couche de litière n'est pas uniforme en épaisseur. Nos calculs devront tenir compte de cette hétérogénéité en considérant par exemple, l'émissivité du système sol + litière comme l'émissivité moyenne calculée pour différentes épaisseurs de la litière